\documentstyle[12pt,epsfig,amssymb,amsmath]{article}
\def\beqa{\begin{eqnarray}}
\def\eqar{\end{array}}
\def\beqar{\begin{array}}
\def\eqa{\end{eqnarray}}
\def\bars{\begin{eqnarray*}}
\def\ears{\end{eqnarray*}}
\def\ov{\overline}

\def\beqa{\begin{eqnarray}}
\def\eqar{\end{array}}
\def\beqar{\begin{array}}
\def\eqa{\end{eqnarray}}
\def\bars{\begin{eqnarray*}}
\def\ears{\end{eqnarray*}}
\def\ov{\overline}

\newcommand\lr[1]{{\left({#1}\right)}}

\newcommand{\bt}{\begin{tabular}}
\newcommand{\et}{\end{tabular}}
\newcommand{\bd}{\begin{displaymath}}
\newcommand{\ed}{\end{displaymath}\noindent}
\newcommand{\ec}{\end{center}}
\newcommand{\bc}{\begin{center}}

\newcommand{\x}{x_{b_j}}

\begin{document}
\title{Conformal couplings and ``azimuthal matching'' of QCD Pomerons}
\author{N. Marchal\thanks{Stagiaire de Dipl\^ome d'Etudes Approfondies de Physique Th\'eorique (ENS,Paris)} \ and R. Peschanski\\
Service de Physique Th\'eorique  CEA-Saclay \\ F-91191
Gif-sur-Yvette Cedex, France\\
}
\maketitle

\begin{abstract}

Using the asymptotic conformal invariance of perturbative QCD  we derive the expression of  the coupling of external states to all conformal spin $p$ components of the forward elastic amplitude. Using the wave-function formalism for structure functions at small $x,$  we derive the perturbative coupling of  the virtual photon for $p= 1$ , which is maximal for linear transverse polarization. The non-perturbative coupling to  the proton is discussed in terms of ``azimuthal matching''  between the proton color dipoles and
the $q\bar q$ configurations of the photon. As an application, the recent conjecture of a second QCD Pomeron related to the conformal spin-1 component is shown to rely upon  a strong  azimuthal  matching of the $p= 1$ component
in $\gamma^*$-proton scattering.
\end{abstract}

\section{Conformal invariance of the BFKL equation}

As well-known the Balitskii, Fadin, Kuraev, Lipatov (BFKL) equation \cite {bfkl}
 expresses
the elastic amplitude of two off-shell gluons in the high energy limit corresponding to the perturbative QCD resummation of leading logarithms. In terms of transverse coordinates (Fourier transforms of the four external gluon transverse momenta), the equation can be schematically written 
$\frac{\partial f}{\partial Y} \lr{k, k',q;Y} =  K \otimes  f,$
where $Y$ (in the case of  structure functions $Y=\log 1/\x$) is the whole rapidity range, $k,k',$ the two-dimensional initial gluon momenta and $q,$ the  2-momentum transfer. The BFKL integro-differential kernel $K$ is known to possess a global conformal $SL\lr{2, \mathbb{C}}$ invariance
\cite{Li86}.  The BFKL derivation is made in the framework of the leading log approximation but it is interesting to investigate the more general consequences of the asymptotic conformal invariance, which could be maintained at higher order. Indeed, for instance, next leading BFKL calculations could preserve an approximate conformal invariance \cite {brod}. Deviations from asymptotic conformal invariance could also be studied by comparison with the results obtained with this assumption.

The solution of the BFKL equation is for the 4-point gluon amplitude. For practical application to the proton structure functions, say,  the conformal couplings
of the BFKL solution with the $q\bar q$ states of the virtual photon and with the proton have to be explicited. This is the main purpose of our paper to formulate the most general coupling of external states and discuss the constraints imposed by the conformal symmetry  of the BFKL equation.

The conformal symmetry of the BFKL equation \cite{Li86} is a 
powerful tool. Knowing that  the kernel $K$ is invariant in the $SL\lr{2, \mathbb{C}}$ transformations, it is possible \cite{Li86,exact} to solve exactly the BFKL equation by expanding over the $SL\lr{2, \mathbb{C}}$ unitary irreducible  representations which are labelled by two quantum numbers, namely the ``conformal dimension'' $\gamma = \frac{1}{2} + i\nu$ and the ``conformal spin''\footnote{The conformal spin can be half-integer but only integer values contribute to the structure functions \cite {nawa}. It can also in principle take negative half-integer and integer values \cite {Li86}, but the decomposition over positive eigenvalues is complete and thus sufficient to describe the conformal expansion.} $p \in \Bbb{Z}.$ In the appropriate eigenbasis 
$K$ is diagonal with eigenvalues
\begin{equation}
\epsilon (p ,\gamma) =  \bar \alpha \ \chi_{p} (\gamma)
\label{omega}
\end{equation}
where $\bar \alpha = \frac {\alpha N_c}{\pi}$ and 
\beqa
\chi_p (\gamma) & = &  2 \Psi (1) 
-\Psi (p+1-\gamma) - \Psi (p+\gamma)\\
& = & \sum_{m=0}^{\infty} \left\{ \frac{1}{m+\gamma+p} +\frac{1}{p+1-\gamma +m} -\frac{2}{m+1} \right\}.
\label{chi}
\eqa
Using the expansion over the whole conformal basis leads to an expression for the structure function as
\begin{equation}
F_2 (Y, Q^2) = \ \sum_{p} F_p (Y, Q^2)\equiv \ \sum_{p}\int_{\frac 12-i\infty}^{\frac 12+i\infty} d\gamma \lr{ \frac{Q}{Q_0}}^{2\gamma} e^{\ov{\alpha} \chi_p (\gamma)Y} f_p (\gamma)
\label{eq02}
\end{equation}
where $f_p (\gamma)$ is obtained from the 
couplings of the different conformal spin components to the external sources.
The aim of our paper is to discuss these functions $f_p (\gamma)$ taking into account the constraints due to  conformal invariance.

In the expression (\ref{eq02}), one usually sticks to the component $p=0$
which gives rise to the ``hard'' QCD Pomeron in the leading order
BFKL formalism. In phenomenological applications, the perturbative coupling of the conformal component $p=0$ to the virtual photon has been known since a long time \cite {bjorken,kT,nikzak,charme} and some models of the non-perturbative coupling to the proton have been discussed \cite {charme}. 

However, little has been done about  higher conformal spins. They have been considered in  two-jet production with large rapidity interval in hadron-hadron collisions \cite {del1} and in the forward jet production in Deep Inelastic Scattering \cite {del2} which correspond to  two ``hard'' vertices with similar characteristic scales. We shall come back to the corresponding perturbative QCD calculations later on in the discussion. More recently, the  general conformal coupling  has
 been formally derived in the eikonal approximation \cite {vertex}, leading to interesting selection rules. But higher spin components were expected to have no practical applications at high energy (rapidity interval) since they are at first sight power suppressed  in energy. It is indeed the case for  the  processes considered in Refs. \cite {del1,del2}. 

However, recently, it has been noticed \cite {pomerons}
that the spin component $p= 1$ may have a non negligeable impact for  processes corresponding to vertices with  different characteristic scales and in particular for proton structure functions at moderate and large $Q^2.$ This is due to a ``sliding'' mechanism which shifts up its effective intercept and thus drastically changes  the energy dependence. The    $p= 1$ spin component may even be interpreted as the remnant of  the well-known ``soft'' Pomeron in the high $Q^2$ region. This result is to be put in perspective with the two-Pomeron conjecture of Ref. \cite {dolan}, where the  ``soft'' Pomeron
 is considered to be higher-twist, while the ``hard''  Pomeron would represent some kind of leading-twist\footnote{Note however that the paper \cite {dolan} is written in the conventional Regge formalism while the study of \cite {pomerons} is made in the framework of the BFKL equation and its conformal invariant setting.}. Hence, it is worth  studying in detail the constraints and properties of conformal couplings to QCD Pomerons, both from a perturbative (for the virtual photon) and non-perturbative (for the proton) points of view. 

The next section {\bf 2} is devoted to the general formalism for the coupling to a generic conformal spin component of the BFKL solution. In section {\bf 3}, we derive the perturbative coupling to the virtual photon wave-function in terms of its $q\bar q$ configurations and introduce a class of models for the non-perturbative couplings to the proton satisfying appropriate constraints. Then, in section {\bf 4}, we make a phenomenological application to the two-pomeron conjecture based on conformal spin components of the proton structure functions, which leads to the necessity of a strong azimuthal ``matching'' condition which is discussed in detail. Summary and conclusions are presented in {\bf 5}.
 
\section{Conformal impact factors}

Following Ref. \cite {Li86} the virtual photon-proton elastic BFKL  scattering amplitude reads
\begin{equation}
{A}(s,-q^2)= i s \int\:\frac{d\omega}{2i\pi} \left( \frac{s}{Q^2} \right)^{\omega} f_{\omega}(q^2)\ ,
\label{eq3}
\end{equation} 
where $s/Q^2\approx 1/x,$ $q^2$ is the  quadri momentum transfer squared and
\begin{equation}
 f_{\omega}(q^2) = \int \: d^2k\: d^2k' \: {\cal V}^{(1)}(k,q)\: \overline {\cal V}^{(2)}(k',q)\:
f_{\omega}(k,k',q)\ .
\end{equation}
$f_{\omega} (k,k',q)$ is nothing else than the $Y \!\to \!\omega$ Mellin transformed of the two-gluon elastic amplitude verifying the BFKL evolution equation (see section {\bf 1}). ${\cal V}^{(1)}(k,q)$ and ${\cal V}^{(2)}(k',q)$ are the so-called {\it impact factors} describing the coupling of the initial states to the gluons.

After straightforward calculations using the conformal basis of eigenvectors
\cite {Li86,exact}, one may write
\begin{equation}
f_{\omega}(q^2)= \sum_p \int \: \frac {d\gamma}{2i\pi} \frac{c(p,\gamma)}{\omega \!- \!\epsilon (p ,\gamma)} \: V_1^{p,\gamma} (q)\ \overline{V}_2^{p,\gamma} (q)\ ,
\label{eq6}
\end{equation} 
with
\begin{equation}
V_{1,2}^{p,\gamma} (q) =\frac{1}{(2\pi)^3} \int \: d^2 \rho \:d^2 \rho' \: d^2k \: {\cal V}^{(1,2)} (k,q)\  e^{ik\rho+i(q-k)\rho'} \: E^{p,\gamma}(\rho, \rho'),
\label{eq2}
\end{equation}
where $\epsilon (p ,\gamma)$ is given in  (\ref{omega}) and $E^{p,\gamma}(\rho, \rho')$  are the   $SL\lr{2, \mathbb{C}}$  eigenfunctions 
\begin{equation}
E^{p,\gamma}(\rho, \rho') =\left(\frac{\rho-\rho'}{\rho \rho'}\right)^{\gamma-p}\ \left(\frac{\bar \rho-\bar {\rho}'}{\bar {\rho} \bar {\rho}'}\right)^{\gamma+p},  \label{10}
\end{equation}
and
\begin{equation}
c(p,\gamma)= \frac { \nu^2 + p^2 }{ \left( \nu^2 + (p- \frac1{2})^2\right) \left( \nu^2 + (p+ \frac {1}{2})^2  \right)}\ ,
\end{equation}
where $\gamma \equiv \frac 12 + i\nu.$
In the forward direction ($q= 0$)  the formula (\ref {eq2}) simplifies. After changing variables to  $\rho+\rho'=2b$ and $\rho-\rho'=r$  and  combining the relations (see ~\cite{Li86})
\begin{equation}
\int d^2b \: E^{p,\gamma} (b\!+\!\frac{r}{2}, b\!-\!\frac{r}{2})
= \frac{b_{p,\gamma}}{(2 \pi )^2} \ r ^{\gamma-p} \ \bar r ^{\gamma+p} 
\end{equation}
and (see ~\cite{gr}) 
\begin{equation}
 \int d^2u \: u^{\gamma-p} \: \overline{u}^{\gamma+p} \: e^{\frac i2 (u+\bar u)} = 2\pi \int d| u| \: |u|^{1+2\gamma} J_{2p} (|u|) = 4^{1+\gamma}\pi  \ \frac {\Gamma(\gamma+p+1)}{\Gamma(p-\gamma)}\ ,
\end{equation}
 one gets
\begin{equation}
V_{1,2}^{p,\gamma} (q\!=\!0) = 2^{1+2\gamma} \ \frac {\Gamma(\gamma+p+1)}{\Gamma(p-\gamma)}
\ b_{p,\gamma}
 \int \: d^2 k \: {\cal V}^{(1,2)} (k) \: k^{-(\gamma+p+1)} \overline{k}^{-(\gamma-p+1)}\ ,
\end{equation}
where
$b_{p,\gamma}$ is a  $SL\lr{2, \mathbb{C}}$ constant  given in Ref.\cite{Li86} and verifying  $ \left| b_{p,\gamma} \right|^2  = \frac{ \pi^6 }{p^2 + \nu^2}.$

Using the relation $\Im m{A}(q^2=0) \equiv s \sigma_{tot} = \frac{s}{4\pi Q^2} F_2(Y,Q^2),$ one  finally obtains
\begin{multline}
F_2 (x, Q^2) \sim \sum_{p} \int d\gamma \ x^{-\epsilon (p ,\gamma)} \left( \frac{Q}{Q_0} \right)^{2\gamma} \ V_1 \ \overline{V}_2\\
= \sum_{p} \int d\gamma \:\left| \frac{ \Gamma \left( p+\gamma \right)}{ \Gamma \left( p - \gamma +1 \right) } \right|^2  x^{-\epsilon (p,\gamma)} \left( \frac{Q}{Q_0} \right)^{2\gamma} \\
\times \int d^2 \kappa \: {\cal V}_1 (\kappa) \: \kappa^{-(\gamma+p+1)} \: \overline{\kappa}^{-(\gamma-p+1)} 
 \int d^2 \kappa_0 \: \overline{{\cal V}_2} (\kappa_0)  \: \kappa_0^{-2+\gamma +p} \: \overline{\kappa_0}^{-2+\gamma-p}\ ,
\label{vertices}
\end{multline}
where one introduces the natural scaling  variables $ k /Q = \kappa$ for the (photon) vertex $V_1$ and    $ k_0/Q_0 = \kappa_0 $ for the (proton) vertex $V_2.$ Note that the Gamma function prefactors boil down to a factor $1$ on the
integration line over the imaginary axis $\gamma = \frac 12 +i\nu.$

Let us consider for instance the  first components ($p=0,  1$). By separation of modulus and azimuthal integration over $\kappa,$  they correspond to the two first coefficients of the Fourier expansion
\begin{equation}
{\cal V}_{1,2} (\kappa)  =  \alpha_{1,2} (\left| \kappa \right| ) + \beta_{1,2}  (\left| \kappa \right| ) \cos (2\varphi ) + ...
\label{components}
\end{equation}

In the case of proton structure functions and specializing to  the two first components,
one obtains 
\begin{multline}
F_2 (x,Q^2) \sim  \int d\gamma \ x^{-2 \bar \alpha \left( \Psi (1) -{\rm Re} \Psi (\gamma ) \right) } \left( \frac{Q}{Q_0} \right)^{2\gamma} f_0 (\gamma) \\
 +\int d\gamma \ x^{-2 \bar \alpha \left( \Psi (1) - {\rm Re} \Psi (\gamma +1) \right) } \left( \frac{Q}{Q_0} \right)^{2\gamma} f_1 (\gamma) \\ 
 + \sum_{p \neq 0,1,-1}\int d\gamma \ldots
\label{eqf2}
\end{multline}
with
\begin{equation}
f_0 (\gamma ) =  \int_0^{\infty} d\left| \kappa \right| \left| \kappa \right|^{-1-2\gamma}  \alpha_1(|\kappa| )   \int_0^{\infty} d\left| \kappa_0 \right|    \left| \kappa_0 \right|^{-3+2\gamma}\alpha_2(|\kappa_0| )
\label{f0}
\end{equation}
\begin{equation}
f_1 (\gamma ) =  \int_0^{\infty} d\left| \kappa \right| \left| \kappa \right|^{-1-2\gamma} \beta_1(|\kappa| ) \int_0^{\infty} d\left| \kappa_0\right|     \left| \kappa_0 \right|^{-3+2\gamma}\beta_2(|\kappa_0| )\ .
\label{f1}
\end{equation}
Note a positivity constraint in the case of the eikonal coupling for which ~\cite{vertex}
\[ {\cal V} (\kappa) \propto 4 \int d^2r \ \Phi (r)\ \sin^2 \left(  \kappa r/2 \right) \]
where $\Phi (r)$ is the probability distribution of the $q\bar q$ configurations in coordinate space . Hence, 
a positivity condition ${\cal V} (\kappa)>0$ holds which  leads to $\left| \beta \right|< \alpha.$ However $\beta$ can be negative as it is indeed the case in some processes like forward jet production in DIS \cite {del2}. Note also that the positivity constraint does not hold if there are not only $q\bar q$ configurations in the Fock space of the target (e.g. for the proton). 

\section{Conformal couplings to $q\bar q$ configurations}

Let us first derive the conformal couplings to the virtual photon. In the perturbative QCD framework and for the $p=0$ component, it is possible to derive  the couplings from
first order (virtual) gluon- (virtual) photon fusion graphs, thanks to the $k_T$-factorization property \cite {kT}.  Our aim is to start from these results and derive the corresponding coupling to higher spin components. In fact, for the simple reason of the spin $1$ of the virtual photon,  only the  conformal spin $p= 1$ can be obtained from the   transverse polarization components of the photon.

Interestingly, the factorization properties of QCD in the high-energy regime can be put into two equivalent forms \cite {charme}. As sketched in Fig.1, the perturbative\footnote{In the non-perturbative regime, some modifications of the discussion have to be introduced~\cite{charme} due to the fact that the intermediate gluon $g^*$ may be soft enough to be included in the non-perturbative input. In that case the two pictures lead to two different parametrizations.} coupling of the virtual photon to a dipole can be described by two different factorized formulae. One way is to use the $k_T$-factorization property \cite {kT} which relates the $\gamma^*$-dipole cross-section to the product of the impact factors $V$ by a $g^*$-dipole cross-section where $g^*$ is  an off-mass-shell gluon. Another equivalent way is to use the photon wave-function formalism~\cite{bjorken} which uses the $q\bar q$-dipole cross-section where the  $q\bar q$ configurations are defined by the virtual photon wave function. The target dipole is considered to be small (or massive) in order to justify the (resummed) perturbative QCD calculations.

We shall thus make use of the relation (see Fig.1 and Ref.~\cite{charme}) between the impact factors  and  the wave-functions~\cite{bjorken} of the transverse photon in terms of  $q\bar q$ configurations for  both helicities. This relation reads~\cite{charme} for $p=0$ 
\begin{equation}
\phi_{T}^{(p=0)}(\gamma)\equiv \frac 1{2\pi}\int  rdrd\varphi \lr{r^2Q^2}^{1-\gamma} \int dz 
\lr{\vert\Psi^+_{T}\lr{r,z}\vert^2+\vert\Psi^-_{T}\lr{r,z}\vert^2}
= C\  \frac {V_{T}^{(p=0)}}{\gamma}\ \frac {1} {v(1\!-\!\gamma)} \ ,
\label{relation}
\end{equation}
where 
\begin{equation}
v(1\!-\!\gamma) \equiv
2^{2\gamma-3}\frac{\Gamma (1\!+\!\gamma)}
{\gamma (1\!-\!\gamma)\Gamma(2\!-\!\gamma)}.
\label{gluon}
\end{equation}
$v(\gamma)$ is   the factorized coupling of the off-mass-shell gluon to a dipole~\cite{charme}. The light-cone wave functions of the transverse photon $\Psi^+_{T}$ for helicity $^+$ and $\Psi^-_{T}$ for helicity $^-$ are ~\cite{bjorken} 
\begin{eqnarray}
\Psi^+_T \lr{z,r,Q^2}  = \sqrt C\  z\ e^{i\varphi} \epsilon  K_1\lr{\epsilon r} \\
\Psi^-_T \lr{z,r,Q^2}  = \sqrt C\   \lr{1- z}\ e^{-i\varphi} \epsilon  K_1\lr{\epsilon r} ,
\label{waves}
\end{eqnarray}
where $K_1$ is the Bessel function. By definition $\epsilon \equiv Q\sqrt{z(1-z)}$ and  the normalization is
$C= \frac{\alpha_{em}N_c e^2}{4\pi\alpha_s}.$

Now, for an arbitrary combination of both helicities, one finds contributions to two Fourier components in azimuthal angle (see (\ref{components})), namely
\begin{multline}
\left| \eta_+ \Psi_T^+ + \eta_- \Psi_T^- \right|^2 \sim \left| \eta_+ z e^{i\varphi} + \eta_- (1-z) e^{-i\varphi} \right|^2 \epsilon^2  K_1^2\lr{\epsilon r} \\= \left\{\underbrace{\eta_+^2 z^2 +\eta_-^2 (1-z)^2}_{(p=0)} + \underbrace{2\eta_+ \eta_- z (1-z)}_{(p= 1)}\ \cos 2\varphi\right\}\  \epsilon ^2  K_1^2\lr{\epsilon r}\ .
\label{combine}
\end{multline}
Normalizing to $\eta_+^2 + \eta_-^2 =1,$  It is easy to realize that the two linearly polarized components $ \eta_+ = \pm \eta_- = \frac{1}{\sqrt{2}}$ give opposite contributions\footnote {Note an overall sign ambiguity, which has to be fixed by the calculation of both vertices in the process. For instance the overall sign is negative in the forward jet case \cite {del2}.} to  the component $p= 1.$ The coupling to the linearly polarized photon are  obtained by inserting the appropriate $z$-dependent factor in the expression of the wave function contribution to the $p= 1$ component. Projecting on the $p=1$ azimuthal Fourier component, one writes
\begin{multline}
\phi_T^{(p=1)}(\gamma)\equiv\frac{1}{2\pi}\int  \ 2\cos\varphi\ rdrd\varphi\lr{r^2Q^2}^{1-\gamma} \int dz\: 2\Re e\lr{\Psi^+_{T}\ {\Psi^-_{T}}^*}\lr{r,z}
\nonumber\\
= \frac{\alpha_{em}N_c e^2}{4\pi\alpha_s} \int rdr \lr{r^2Q^2}^{1-\gamma} \int dz \:  2z(1-z)\epsilon^2 K_1^2 (\epsilon r)
\nonumber\\
  \sim  \int du \: u^{3-2\gamma} \: K_1^2 (u)\  \times \ \int dz \:  2z^{\gamma} (1-z)^{\gamma},
\end{multline}
or, noting that the only difference between the two components come from the $z$-dependent factors,
\begin{equation}
\frac {\phi_T^{{(p=1)}}}{\phi_T^{(p=0)}} = \frac {V_T^{{(p=1)}}}{V_T^{(p=0)}}\equiv \ \frac{\gamma}{\gamma + 1}.
\label{p1}
\end{equation}
One finally gets:
\bars
 \lr{ \beqar{c} \phi_T^{(p=0)} \\ \phi_T^{{(p=1)}} \eqar } &\sim& \int \frac{d^2r}{2\pi} \lr{r^2Q^2}^{1-\gamma} \int dz \: \lr{ \beqar{c} z^2+(1-z)^2\\ 2z(1-z)\eqar } \epsilon ^2 K_1^2 (\epsilon r)
\\
 & = &   2^{1\!-\!2\gamma}
\frac{\Gamma^2(1\!+\!\gamma)\Gamma^2(1\!-\!\gamma)\Gamma(2\!-\!\gamma)\Gamma(3\!-\!\gamma)}{\Gamma(2\!+\!2\gamma)\Gamma(4\!-\!2\gamma)}\: \lr{ \beqar{c} \frac {1+\gamma}{\gamma}\\ 1\eqar }\ .
\label{phis}
\ears
Using (\ref {p1}) and the relation (\ref {relation}), it is easy to write down
the similar relation for the impact factors $V_T^{(p=0,1)}.$

As we just saw, the perturbative photon couplings to the non zero conformal spins requires a non zero linear polarization of the $q\bar q$ wave-function of the transverse photon to be dynamically active in the reaction. In other terms, the $p=  1$
BFKL component requires a maximal azimuthal correlation while the $p=0$ one is
completely decorrelated azimuthally. Partial azimuthal (de)correlation can be obtained by a mixture of different BFKL components. This will in general depend on the 
dynamical features of the overall reaction. For instance \cite {del2}, forward jet production in DIS can lead to some azimuthal correlation at small rapidity interval where the higher spin component  $p=  1$ are still present. However the general prediction is a significant azimuthal decorrelation due to the strong 
dominance of the $p=0$ component in this case.

In the case of proton structure functions, however, the ``sliding mechanism'' is able \cite {pomerons} to promote the 
higher spin components, especially for $p= 1,$ to be still important at high energy (and relatively low $Q^2$) and thus to keep  rather strong azimuthal correlations present in that region. This implies a discussion of the non-perturbative couplings. An important remark is that the ``sliding mechanism'' is also expected for  perturbative couplings when a large ratio exists between   the characteristic scales of both vertices. It would thus  also be interesting to study such processes where we would  predict an increase of the azimuthal correlations accompanying the
expected ``sliding mechanism''.

The non-perturbative couplings, e.g. to the proton, are in general beyond our
present theoretical knowledge. It is already true for the leading $p=0$ conformal components, where there are some ambiguities~\cite{charme} in the way one is able to factorize the perturbative from the non-perturbative couplings. This is all the more true for the non-leading $p= 1$ component which, to our knowledge, are for the first time studied for proton structure functions in the present paper. For the sake of definiteness, we will follow some reasonable theoretical and phenomenological requirements which we now indicate:

i) The interaction of the proton are governed (at small $x$) by its $q\bar q$ configurations. This can be interpreted as color dipole configurations \cite {dipoles,charme}. Compared with those of the virtual photon, their quantum fluctuations around the proton size $Q_0$  are expected to be smaller.

ii) The coupling of the proton will be asked to obey the ``sliding mechanism'', that is to verify the convergence and analyticity properties found in Ref.\cite{pomerons}. In particular, no singularity with $\gamma > -1$ should appear in the $p=1$ coupling.

iii) Within conditions i) and ii), the $p=0$  ($\alpha_2$ in (\ref{components})) and $p=1$ ($\beta_2$ in (\ref{components})) couplings will be assumed to be equal up to a normalization which will be determined phenomenologically and  represent the necessary degree of {\it azimuthal correlation} for practical relevance. 

iv) Concerning the abovementionned sign ambiguity of the $p=1$ vertices, it is removed for 
the contribution to structure functions which ought to be positive. Thus the product of the photon and proton vertices is considered to be positive.

We shall now propose a convenient class of parametrizations of the proton couplings $\alpha_2, \beta_2$ in equations (\ref {vertices},\ref {components}) satisfying the requirements i)-iv). Noting \cite{gr} the relation 
\begin{equation}
\int_0^{\infty} d\kappa_0\: \frac{\kappa_0 ^{q-1+2\gamma}}{1+ \kappa_0 ^{2q}} \equiv \frac1 {2q}\ B\left(\frac 12 \!+\!\frac {\gamma}q, \frac 12 \!-\!\frac {\gamma}q\right)
=\frac {\pi}{2q \cos\left(\frac {\gamma}q\right)}\ ,
\label{grad}
\end{equation}
 we  are led to choose 
\begin{equation}
\alpha_2, \beta_2 \propto \frac{ \kappa_0 ^\lr{q+2}}{1+\kappa_0 ^{2q}} ;\  f_0(\gamma), f_1 (\gamma) \propto \frac 1{ \cos\left(\frac {\gamma}q\right)}\ .
\label{models}
\end{equation}
 Eventually, one may  vary the peaking of the distribution around $\kappa_0 =1$ by changing the values of $q.$ It is interesting to note that for $q \ge 2$ the gauge invariance constraint \cite{Li86} $\alpha_2(0)= \beta_2(0)=0  $ is automatically verified. One can also multipoly by a polynomial expression in $\gamma.$ This can be used to satisfy  the constraints, in particular the analyticity ones by cancelling poles at $\gamma>-1.$

\section{The two-Pomeron conjecture and azimuthal matching}

As already mentionned, the non-zero conformal spin components are generally neglected in the phenomenology related to the BFKL equation. Indeed, at ultra-high energy 
$Y\to\infty,$ the structure function components in formula (\ref{eq02}) are
driven by the saddle-points at $\gamma = \frac 12.$ It is easy to realize that the corresponding intercepts
$\chi_p(\frac 12)$ are all negative  for $p \ne 0.$ In the same time their  effective anomalous dimension $\frac 12$ means that they all contribute to a leading-twist behaviour. However, it has been remarked in Ref. \cite{pomerons} that at large but finite values of $Y$ or $Q^2$ the corresponding saddle-points  {\it slide}  away from $\gamma = \frac 12$ and generate contributions with very different $Y$ and $Q^2$ behaviour from the 
ultra-asymptotic ones. In particular the $p =1$ component is still increasing with energy (positive intercept) and their $Q^2$ behaviour  mimic an higher-twist behaviour, i.e. they decrease like a negative power of $Q^2.$ Both features  allowed the authors of \cite{pomerons} to look for the possibility that the  $p = 1$ component
could be interpreted as the high $Q^2$ remnant of the ``soft'' pomeron considered as an higher-twist contribution from the point of view of the operator product expansion of QCD. This would provide a QCD framework for the 
two-Pomeron hypothesis proposed in \cite {dolan}  to describe 
 the phenomenological  features of structure functions  in a  different, Regge approach.

Let us now investigate how the phenomenological discussion can be influenced 
by the determination of the conformal couplings derived in the previous sections. In order to analyze the phenomenology of structure functions in a manner similar to Refs. \cite {pomerons,dolan}, we have to introduce  our determination (\ref{phis}) of the perturbative coupling to the photon and discuss the  proton coupling using, for instance, the family of parametrizations  (\ref{models}). In the discussion, however, it is important to 
 take into account the ambiguity of the separation between perturbative and non-perturbative couplings discussed  in \cite {charme} for the $p=0$ component. Let us recall the problem and extend its lessons to the $p= 1$ component.

We will consider the  following parametrization\footnote{In \cite {charme}, two different models were introduced, depending whether the factorization between perturbative and non-perturbative couplings is assumed at the intermediate gluon level (model I in \cite {charme}) or at the quark level  (model II in \cite {charme}). This ambiguity relies on the 
possibility of the gluon coupling to the $q\bar q$ configurations of the photon (with its typical singularity in $1/\gamma$) to be present (model I) or absorbed (model II)  in the non-perturbative coupling to the proton. We checked that the results we obtained in the framework of model I are very similar for model II up to a renormalization of the $p=1$ component.} of the functions $f_p$ to be inserted in (\ref{eq02}):
\begin{equation}
f_0 (\gamma)  =  \phi^{(p=0)}_T (\gamma) \otimes  \frac{\gamma (\gamma+1)}{ \cos \frac{\pi \gamma}{q} }
\label{I}
\end {equation}
\begin{equation}
f_1 (\gamma)  =  \phi_T^{(p=1)}(\gamma) \otimes {\cal N}_I\ \frac {\gamma (\gamma+1)}{ \cos \frac{\pi \gamma}{q}}, 
\label{I'}
\end {equation}
where the non perturbative coupling has been chosen in order to satisfy the analyticity and convergence constraints in a minimal way.
Assuming the same analytic form for the non-perturbative $p=1$ proton coupling than  $p=0,$ the arbitrary  normalization ${\cal N}_{I}$  quantify 
the relative  weight which we want to evaluate. The value  $q=4$ has been choosen for convenience. $q>2$ at least is needed  to  verify the constraints ii). We checked that the  results are rather independent of these choices, provided the constraints are satisfied.  Note that
$f_1$ is  ``softer'' at $\gamma=0$
than $f_0$ due to the relative factor $\gamma/(\gamma+1).$

On a  physical point of view,  the non-perturbative vertices in formulae (\ref {I},\ref {I'}) can be interpreted \cite {npr,charme} as related to the wave functions of the {\it primordial dipole} configurations in the proton. In the QCD dipole model \cite {dipoles} the BFKL dynamics can be expressed in terms of the dipole-dipole cross-section. Translating this model in   the case of $\gamma^*$-proton scattering, it amounts to consider this cross-section averaged both over the $q\bar q$ configurations of the photon and the {\it primordial dipole} configurations of the proton.

 In order to check\footnote{We used the model I parametrization, but the results are the same for model II or by changing $q>2.$}  the sliding mechanism advocated in \cite{pomerons}, we display 
in Fig.2 the normalization independent plot $\frac{\partial \ln F_p}{\ov{\alpha }\partial Y} $ as a function of  $\frac{\partial \ln F_p}{\partial \ln Q^2}$
for large $Y$ and different values of $\ln Q^2/Q_0^2.$ On the same plot and for the same values is also shown the corresponding results for the Regge parametrization of 
\cite {dolan}. As discussed in  \cite{pomerons}, the results (the black circles in Fig.1)  gives the location in a two-dimensional representation where the effective intercept is plotted as a function of the effective saddle-point $\gamma_c.$ They can be shown \cite {pomerons}  to be situated near the 
curves defined by the functions $\epsilon_p(\gamma),$ independantly of the peculiar form of the factors $f_{0,1}.$ The sizeable sliding  of the $p= 1$ component is proven by the  shift of the corresponding points with respect to the ultra-asymptotic value at $\gamma_c=\frac 12.$ Moreover the 
 evolution at large $ Q^2$ meets the phenomenological determination of the two Pomeron components of \cite {dolan} for $\log \left(Q^2/Q_0^2\right) \sim 8,10$ and reasonable values of the parameter $\bar \alpha \sim .4.$

In order to determine the relative strength of the $p=1$ and $p=0$ components, and thus the r\^ole of the conformal prefactors  $f_{0,1}$
(see formulae (\ref{I} , \ref{I'})) we have considered the 2-Pomeron fit (``hard'' and ``soft'') of \cite {dolan} in the large $Q^2$ region
where it meets\footnote{This comparison is to be taken only with a grain of salt since it is made in a region where the ``soft'' component is weak and thus not directly determined by data. A determination at small $Q^2$ would be more precise
but then, the non-perturbative corrections are expected to be important and may
invalidate a correct evaluation of the normalization in a BFKL framework.} the behaviour  of the two ($p=0$ and $p=1$) conformal spin components. For
instance we show in Fig.3 the $Y$ dependence at fixed large $Q^2.$

The results indicate  large normalizations, namely  ${\cal N}_I \sim 50.$ Some other tests show that the normalization is always large of the same order. Thus, the 2-Pomeron conjecture (as seen from a  QCD point of view) is  obtained only if a strong dynamical 
enhancement favours the non-perturbative coupling of the $p=1$ component. Since the perturbative coupling is  maximal for linearly polarized transverse photon and  limited in size because of positivity constraints (the photon has only $q \bar q$ configurations) the  relevance of the $p=1$ coupling relies on the existence  of a non-perturbative mechanism enhancing considerably the  matching between the proton primordial dipole configurations and the azimuthal polarization of the virtual photon.

We shall now speculate on  such a non-perturbative mechanism   based on azimuthal  matching in $\gamma^*$-proton scattering.

The mechanism  is the following (see Fig.4, for a schematic representation). It is known from a long time \cite {bjorken} that deep-inelastic lepton proton scattering is not necessarily dominated by a 
``hard'' process if the energy is large with respect to the photon virtuality, e.g. if $x \sim Q/W$ is small. Indeed the {\it effective} virtuality is
$\hat Q = Q \sqrt {z(1-z)},$ where $z$ (resp. $1-z$) is the momentum fraction of the quark (resp. antiquark) in the virtual $q\bar q$ state configurations of the virtual photon. This is explicit, for instance, in the 
wave-functions (\ref{waves}). Thus, if the favoured $q\bar q$ configurations are 
particularly assymetric (aligned jet\cite {bjorken} configurations)
one even may reach the situation where the quark or the antiquark in the pair have so small momentum that $\hat Q$ is of order unity and the reaction dominated by a ``soft'' process.

However, the experimental results seem not to favour the   aligned jet mechanism since a ``hard'' component shows up which is precisely the one which
could be described by the $p=0$ component. Yet, for the $p=1$ component, it is not excluded however that a partial jet alignement can take place, at least at moderate $Q^2.$ At high $Q^2$ one could then expect that assymetric configurations are substantially favoured in a kind of ``hard/soft'' compromise:  the effective virtualities $\hat Q$ are smaller than $Q$ while remaining in the ``semi-hard'' regime. As a consequence, one expects a substantial {\it azimuthal matching}  between the $q\bar q$ configurations of the virtual photon and the $q\bar q$ (or primary dipole \cite {npr}) configurations in the proton, see Fig.3.
This {\it azimuthal matching} may give a strong dynamical enhancement for the coupling of the linearly polarized components of the photon to the proton. By this azimuthal enhancement  one could find the qualitative justification for the two-pomeron description to be based on the two conformal spin components. On the other hand, in the absence of such a mechanism
 the $p=1$ components, even if increasing with energy due to the 
sliding phenomenon, would not be coupled enough to the proton to give rise to a sizable component. We shall in conclusion discuss possible tests of azimuthal alignment which is certainly deserving further study.

\section{Summary and Conclusions}

Let us briefly summarize our results:

i) Taking into account that the non-zero (indeed the $p= 1$) conformal spin components of the BFKL QCD Pomeron could be phenomenologically relevant in deep-inelastic lepton proton scattering, we have given the formal expression of the proton structure functions' conformal spin
components in terms of the
appropriate {\it impact factors}.

ii) We have computed the perturbative {\it impact factor} for the $p= 1$ component at the virtual photon vertex using a general relation between impact factors and $q\bar q$ wave functions. the key result is that the coupling is maximal for linear azimuthal polarization and zero for circular (or no) polarization.

iii) In order to be phenomenologically relevant as a ``second'' Pomeron contribution in $\gamma^*$-proton scattering, a strong {\it azimuthal matching} with the primordial dipole $q\bar q$ configurations of the proton is required. This non-perturbative mechanism could be associated in part with jet alignment {\it a la Bjorken} for the $p= 1$ component, while it is expected to be 
weak or absent for the $p=0$ one.

Some comments are in order. The large enhancement (a factor $\sim 50$) of the non-perturbative coupling to the
$p=1$ component that we found necessary to match with the two-Pomeron parametrization of Ref.\cite {dolan} is consistent with the key point of this kind of phenomenological analysis: the mismatch between the ``hard'' and ``soft'' Pomerons is more important than in models with only one effective Pomeron singularity. In particular, at intermediate values of the virtuality $Q^2,$ both contributions are important. This is the reason why the ``hard'' component has a large intercept $\epsilon (p\!=\!0,\frac 12) \simeq 0.4$ in agreement with the theoretical range of values and larger than the effective intercept $\epsilon (p\!=\!0,\frac 12) \leq .3,$ see for instance \cite {npr,charme}.  The validity of a non-negligeable ``soft'' Pomeron coupling at high $Q^2$ is thus to be checked
in further study.

A specific feature of the $p=1$ component is its special azimuthal properties. the question arises whether it is possible to isolate it using azimuthal correlation properties. For the total inclusive process, leading to the determination of the structure function itself, this seems uneasy.  The relevant azimuthal  axis in the photon-dipole center-of-mass frame, see Fig.4,
can be very different from the photon-proton one, and thus, in particular, the  s-channel helicity conservation which seems to be an approximate property of the ``soft'' Pomeron coupling is not in contradiction with the conformal spin properties of the $p=1$ component.

A possible test of the {\it azimuthal matching} could be performed in forward jet production in Deep Inelastic Scattering. Indeed, while the commonly considered configuration with similar scales for the photon probe and the jet
is expected to lead to small azimuthal correlation at high rapidity interval
\cite {del2}, the case with a larger scale ratio is expected to lead to stronger
azimuthal correlation due to the enhancement with energy of the higher conformal spin component responsible for the azimuthal matching in the considered formalism.  Indeed, a practical way of checking the azimuthal correlations could be to fix a certain range of high\footnote {However, in practice  $Q^2$ is limited by the necessity of a large rapidity interval with the forward jet.}
$Q^2$ for the photon virtuality and vary the tranverse momentum of the forward jet down to the lower admissible value to select a jet. By this way, one increases  the ratio $Q^2/k_T^2$ of scales and thus enhance the energy behaviour of  the $p=1$ component. Moreover, since the model implies a strong mismatch between the ``soft'' and ``hard'' Pomerons at intermediate scale, one expects the development of a stronger (and perhaps different in sign!) azimuthal correlation than the perturbatively predicted azimuthal correlations (with negative sign) studied in  ref. \cite {del2}. A similar method can be proposed at Tevatron analyzing azimuthal correlations  between  two jets (1) and (2) in different hemispheres, as analyzed in  ref. \cite {del1}, with the prediction that it will  increase together with the ratio  $k_T^{(1)}/k_T^{(2)}.$ 

\bigskip 

{\bf ACKNOWLEDGEMENTS}

\bigskip 
We are grateful to Stephane Munier and Henri Navelet for stimulating discussions
and suggestions. One of us (N.M.) wishes to thank the ``Service de Physique Th\'eorique de Saclay'' and its staff for the kind hospitality during the period of ``stage''.
\eject

\eject
\noindent
{\bf FIGURE CAPTIONS}
 
\vspace{1.5cm}
\noindent
{\bf Figure 1}
{\it The two factorization schemes of the $\gamma ^*$-dipole cross-section}
\bigskip

The photon ($\gamma ^*$) -dipole ($d$) cross-section, as given by the QCD dipole model  corresponding to the perturbative QCD resummation at small $x$, admits two equivalent factorization schemes (see text). First Scheme: $k_T$-factorization  of the  $g^*$-dipole cross-section with transverse impact factors $V_T^{(p=0,1)}$; Second Scheme: wave-function factorization of the $(q\bar q)-d$ cross-section, where the virtual photon transverse wave functions  $\Psi _T^{(+,-)}$ are described on the basis of its $(q\bar q)$ configurations. The two conformal spin components $(p=0,1)$ of the transverse impact factors can be expressed in terms of the wave functions for left (+) and right (-) helicities, see Eqns. (\ref {p1},\ref {phis}). 

\vspace{0.5cm}
\noindent
{\bf Figure 2} {\it Plot of effective intercept vs. effective dimension at fixed large $Y$}
\bigskip

The effective intercept  $\partial \ln F_{0,1} / \bar \alpha \partial Y$ plotted  vs. the effective anomalous dimension $\partial \ln F_{0,1} / \partial \ln Q^{2}$ is  compared to the   functions $\epsilon_{0,1}\equiv \bar \alpha \chi _{0,1}\left( \gamma \right)$  (see Eqns. (\ref{omega},\ref{chi})). They are computed at $\bar \alpha = .15$ for
  fixed  $Y=10$ and 4 values of $\ln {\displaystyle{Q^{2} / Q_{0}^{2}}}=\left\{ 4,6,8,10\right\} .$  The  weight in the integrals (\ref{eq02}) corresponds to  Eqns. (\ref{I},\ref{I'}).

Black circles:  numerical results;  White circles:  ultra asymptotic saddle points at $\gamma =\frac 12;$ Full lines: the
functions $\epsilon _{p}\left( \gamma \right)$ for $\left(p= 0,1\right);$ Dotted lines; results  from the Regge fit of Ref. \cite{dolan} corresponding to the same value of $Y$ and  $\ln Q^{2}/Q_{0}^{2},$ with $Q_0 \sim 200 {\rm MeV}.$  Arrows indicate the direction of increasing $Q.$

\vspace{0.5cm}   
\noindent
{\bf Figure 3}
{\it  The structure function spin components $F_{2(p=0,1)}$  at fixed large $Q^{2} .$}
\bigskip

The structure function components $F_{2(p=0,1)}$  are displayed  as a function of $Y=\log 1/x$ and compared with the parametrization of the two-Pomeron model of Ref.
\cite {dolan} at $Q^2=1000 {\rm GeV}^2.$ This value is choosen to correspond to $\ln {\displaystyle{Q^{2} / Q_{0}^{2}}}\sim 10 .$

Continuous line: ``hard Pomeron'' component  of Ref.\cite {dolan};
Long-Dashed line: Spin $0$ component.

Short-Dashed line: ``soft Pomeron'' component  of Ref.\cite {dolan};
Dashed line:  Spin $ 1$ component.

\vspace{0.5cm}
\noindent
{\bf Figure 4}
{\it Azimuthal matching of photon and dipole $q\bar q$ configurations}
\bigskip

The photon ($\gamma ^*$) -dipole ($d$) reaction is represented in the center of mass
frame. The azimuthal angle between $q\bar q$ configurations of both colliding systems  is  the angle $\phi$ between the two planes. The quark (resp. antiquark) momentum fraction in the virtual photon is $z$ (resp. $1-z$) (the similar variable for the dipole configurations has been already averaged).

\clearpage

\begin{center}
\epsfig{file=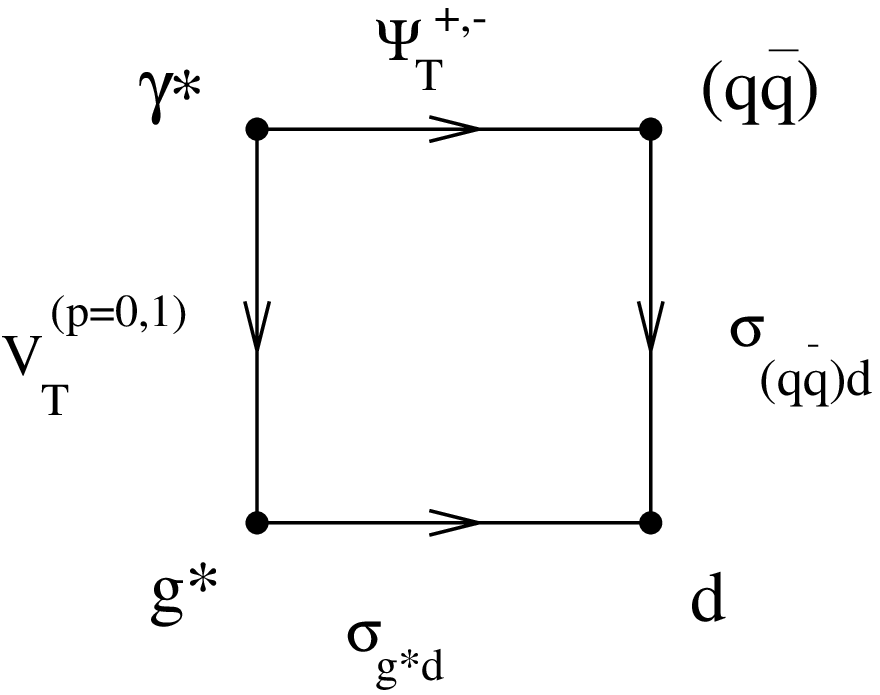,height=10cm}
\end{center}

\begin{center}
{\large\bf Figure 1}
\end{center}

\clearpage

\begin{figure}[h]
\centering
\epsfig{file=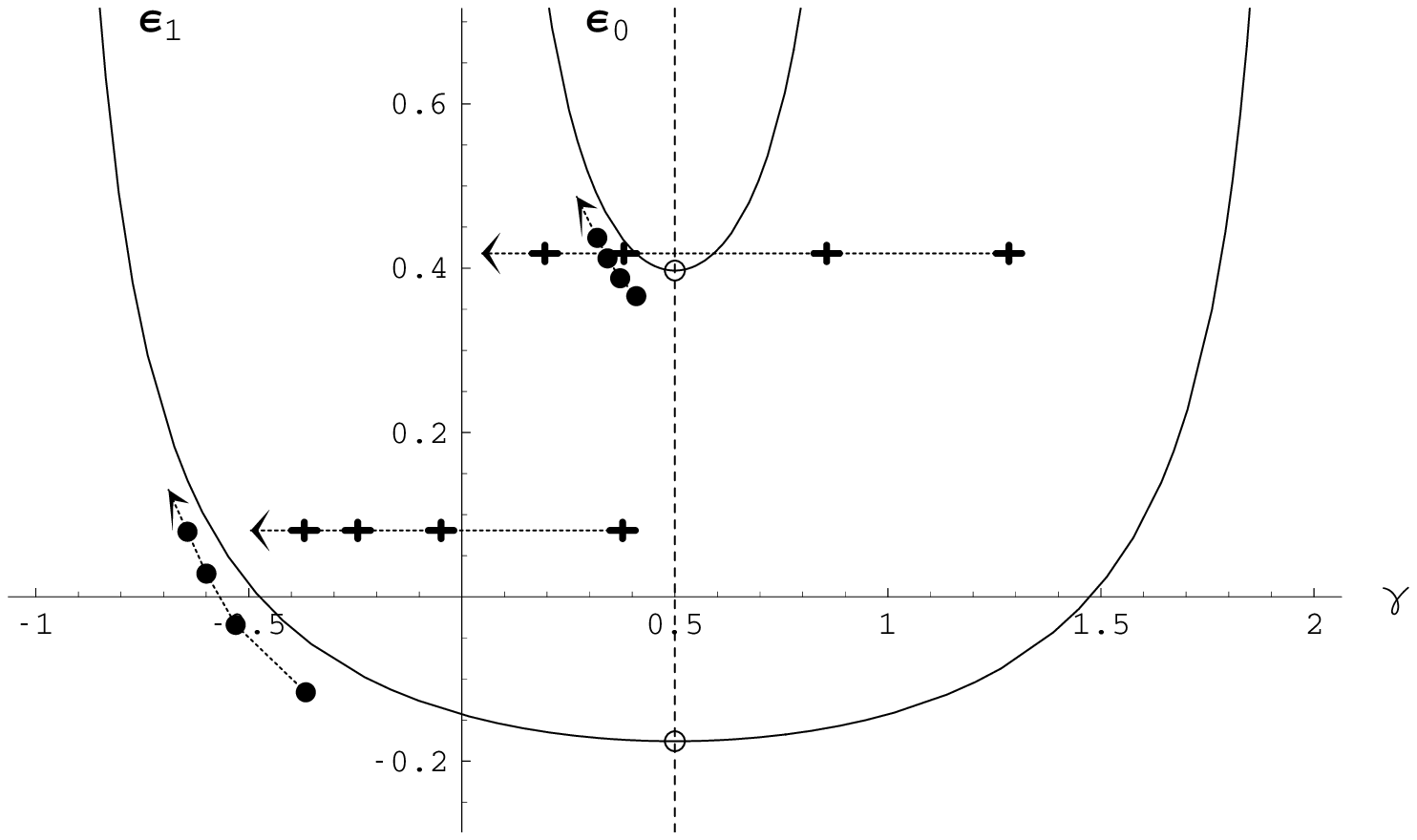, width=15cm}

\begin{center}
\vspace{10cm}
{\Large {\bf Figure 2}}

\end{center}

\end{figure}

\clearpage
\begin{center}
\epsfig{file=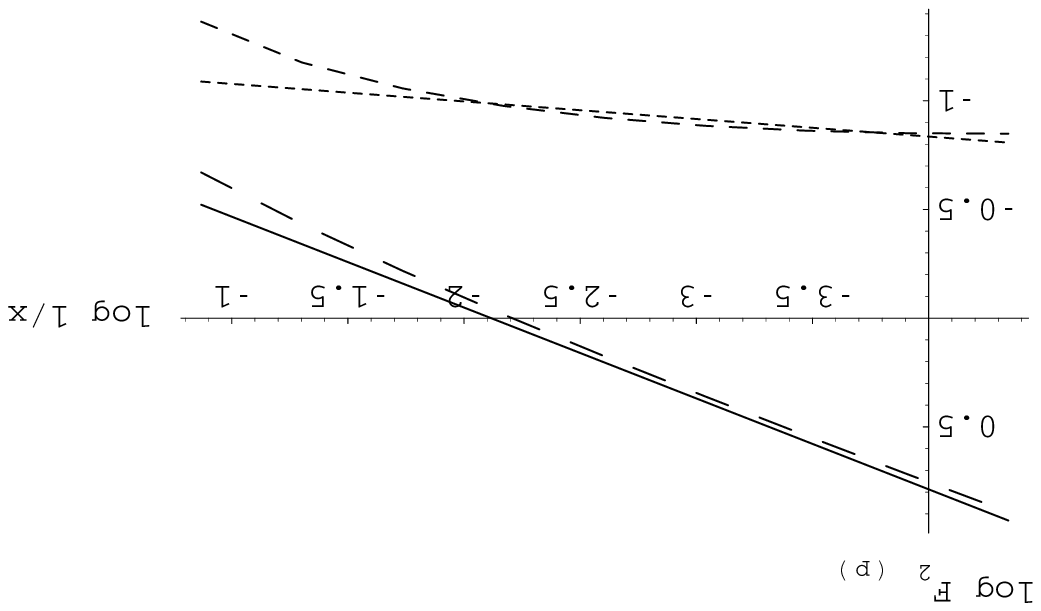,height=10cm,angle=180}
\end{center}

\begin{center}
{\large\bf Figure 3}
\end{center}

\clearpage
\begin{center}
\epsfig{file=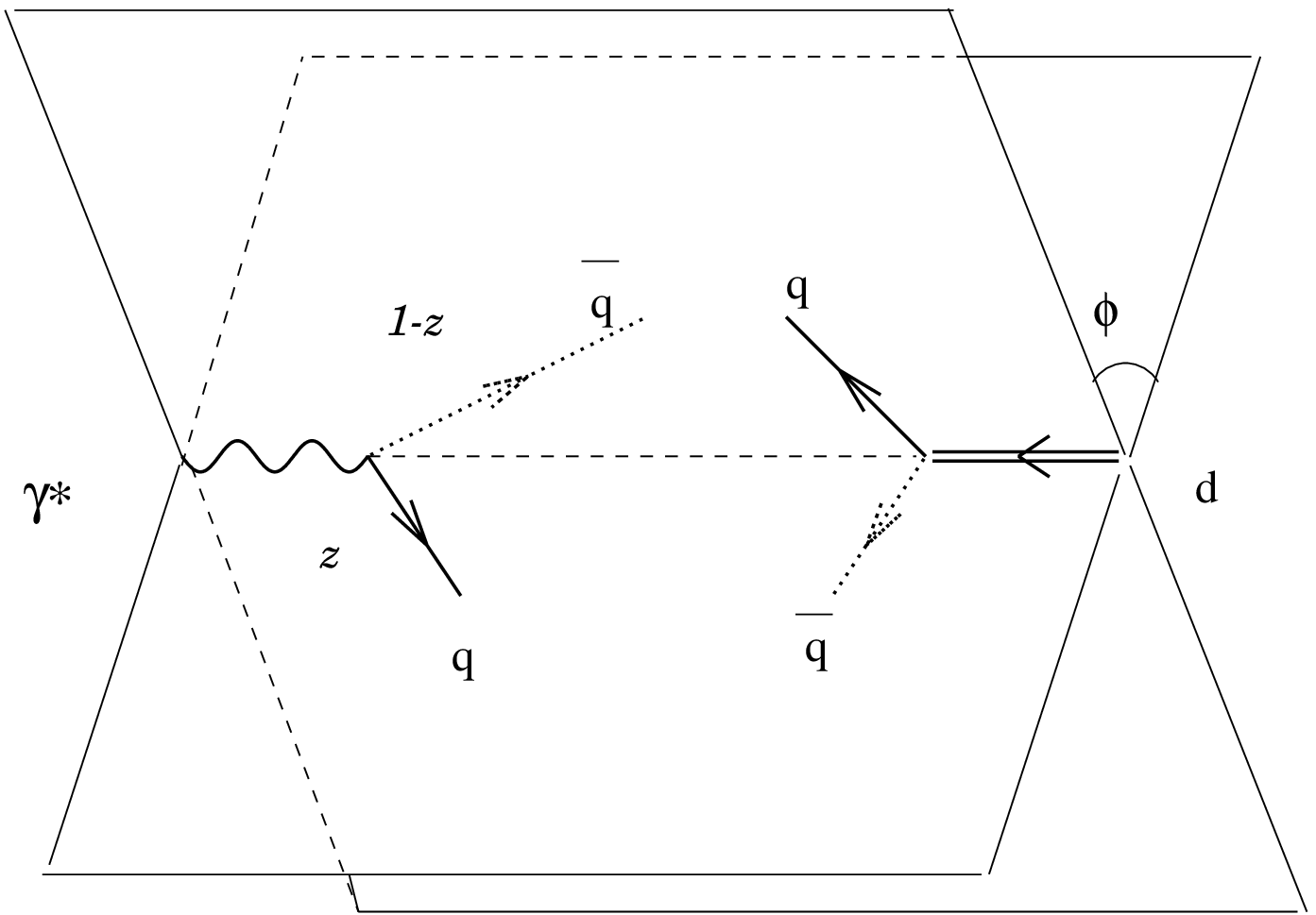,height=10cm}
\end{center}

\begin{center}
{\large\bf Figure 4}
\end{center}

\end{document}